\documentclass[aps,prd,twocolumn,groupedaddress,nofootinbib]{revtex4}

\usepackage{amsmath}
\usepackage{amssymb}
\usepackage{color}
\usepackage{graphicx}

\newcommand{\bk}{{\mathbf k}}

\newcommand{\bq}{{\mathbf q}}
\newcommand{\by}{{\mathbf y}}
\newcommand{\bx}{{\mathbf x}}

\newcommand{\de}{\delta}

\newcommand{\ra}{\rightarrow}
\newcommand{\be}{\begin{equation}}
\newcommand{\ee}{\end{equation}}
\newcommand{\bea}{\begin{eqnarray}}
\newcommand{\eea}{\end{eqnarray}}
\newcommand{\bean}{\begin{eqnarray*}}
\newcommand{\eean}{\end{eqnarray*}}

\newcommand{\vev}[1]{\mbox{$\langle #1 \rangle $}}
\newcommand{\Eqref}[1]{Eq.~(\ref{#1})}

\begin{document}

\title{Gamma-ray observations of blazars \\ and the intergalactic magnetic field spectrum}

\author{C. Caprini}
\affiliation{CNRS, URA 2306 and CEA, IPhT, F-91191 Gif-sur-Yvette, France}

\author{S. Gabici}
\affiliation{APC, Univ Paris Diderot, CNRS/IN2P3, CEA/Irfu, Obs de Paris, Sorbonne Paris Cit\'e, France}

\date{\today}

\begin{abstract}
Very-high energy observations of blazars can be used to constrain the strength of the intergalactic magnetic field. 
A simplifying assumption which is often made is that of a magnetic field of constant strength composed by randomly oriented and identical cells.
In this paper, we demonstrate that a more realistic description of the structure of the intergalactic magnetic field is indeed needed. If such a description is adopted, the observational bounds on the field strength are significantly affected in the limit of short field correlation lengths: in particular, they acquire a dependence on the magnetic field power spectrum. In the case of intergalactic magnetic fields which are generated causally, for which the magnetic field large scale spectral index is $n_B\geq 2$ and even, the observational lower bound becomes more constraining by about a factor 3. If instead $-3<n_B<-2$, the lower bound is significantly relaxed. Such magnetic fields with very red spectra can in principle be produced during inflation, but remain up to now speculative. 
\end{abstract}

\maketitle

\section{Introduction}

Very High Energy (VHE) gamma--rays ($\gtrsim 100$~GeV) from distant extragalactic sources are absorbed by pair production on the intergalactic background radiation field \cite{gould}. Electron--positron pairs produced in this way would then inverse Compton scatter off soft ambient photons, initiating an electromagnetic cascade that might be detected in the multi--GeV to TeV energy domain \cite{pairhalos}. The presence of an InterGalactic Magnetic Field (IGMF) affects the development of the electromagnetic cascade by deflecting (or even isotropizing) electron--positron pairs \cite{pairhalos,plaga}. The deflection would also introduce a delay in the arrival time of the photons produced in the cascade, with respect to the arrival time of the primary VHE photons emitted from the source. For bursting sources, the observation of such a delay would allow to constrain the value of the IGMF \cite{plaga}.

Blazars were the first extragalactic sources to be detected in VHE gamma--rays \cite{michael}, and it was soon realized that they are ideal targets to search for the cascade radiation \cite{pairhalos}. If the IGMF is strong enough to deflect in an appreciable manner the electrons and positrons in the cascade, then such a secondary radiation would appear in the form of a diffuse gamma--ray emission surrounding the central point source of primary photons \cite{pairhalos,Neronov:2006hc}. Moreover, the VHE gamma--ray emission from blazars is characterized by strong, powerful flares \cite[e.g.][]{whippleflare,HEGRA,HESS,MAGIC,VERITAS}. Due to the time delay induced by the IGMF, the cascade radiation would appear as a delayed echo of such flares \cite{dai}.

The IGMF can be modeled as a stochastic magnetic field, statistically homogeneous and isotropic. Both the angular extension $\Theta_{\rm ext}$ and the time delay $t_{\rm delay}$ of the cascade radiation from blazars are expected to depend on the rms strength of the IGMF, but also on its correlation length $\lambda_B$ \cite{neronovsemikoz,durrer}. The IGMF correlation length becomes relevant when it is much smaller than the electron/positron energy loss length for inverse Compton scattering, $\lambda_B\ll D_e$: in this case, the charged particle would undergo several deflections before losing its energy via emission of inverse Compton photons. Its overall deflection angle $\delta$ has been estimated considering that many stochastic deflections would cause the electron/positron to perform a random walk \cite{neronovsemikoz}, leading to 
\be
\delta\simeq\frac{\sqrt{D_e\lambda_B}}{R_L}~~~~ {\rm if}~~\lambda_B\ll D_e\,, \label{deltarandom}
\ee
with $R_L=E_e/(eB)$ the Larmor radius, $E_e$ the electron/positron energy, and $B$ the IGMF strength. On the other hand, if $\lambda_B\gg D_e$, the motion of the charged particle on the energy loss length scale $D_e$ can be approximated as the motion in a homogeneous IGMF, leading to a deflection angle which is simply  \cite{neronovsemikoz}
\be
\delta\simeq\frac{D_e}{R_L}~~~ {\rm if}~~\lambda_B\gg D_e\,. \label{deltahom}
\ee
In the limit of small deflection and neglecting redshift dependences, the angular extension of the cascade radiation is directly proportional to the deflection angle, through $\Theta_{\rm ext}\simeq \delta/\tau$ where $\tau = D/D_{\gamma}$ is the optical depth for the primary gamma rays, $D_\gamma$ denotes the mean free path of the primary photon, and $D$ the distance to the blazar \cite{neronovsemikoz,durrer}. 
The time delay, on the other hand, is given by $t_{\rm delay}\simeq D_\gamma \,\delta^2/2 c$ \cite{dermer,neronovsemikoz,durrer}. It is clear therefore that both $\Theta_{\rm ext}$ and $t_{\rm delay}$ inherit, through $\delta$, a dependence on $B$ and $\lambda_B$: thus, a detection of the cascade radiation would allow to measure a combination of these two (largely unknown!) physical quantities \cite{pairhalos,plaga,neronovsemikoz,coherencescale}. 
Since no cascade radiation has been detected yet \cite{neronovscience,HESShalos} (but see the preprint \cite{buckley}), only constraints on the IGMF can be obtained (see \cite{neronovscience}, or \cite{durrer} and references therein).

For very weak IGMF the angular extension of the cascade radiation is smaller than the instrumental angular resolution, and thus the contribution from the cascade adds up to the primary and point like gamma-ray signal from the blazar. For a given value of the IGMF strength, $B_{\rm PSF}(\lambda_B)$, the extension of the cascade emission would be identical to the instrument angular resolution. Note that, from \Eqref{deltarandom} and from $\Theta_{\rm ext}\simeq \delta/\tau$, $B_{\rm PSF}$ explicitly depends on $\lambda_B$ when $\lambda_B\ll D_e$, as $B_{\rm PSF}\propto \sqrt{D_e/\lambda_B}$ \cite{neronovscience,neronovsemikoz}. For fields significantly stronger than $B_{\rm PSF}(\lambda_B)$, the cascade emission would become quite extended and its contribution to the point like emission would be correspondingly suppressed. 
By using this argument, Neronov and Vovk \cite{neronovscience} obtained a lower limit on the IGMF strength of $B \geq 3 \times 10^{-16}$~G from the non detection by {\it Fermi} of the cascade emission from the TeV blazars 1ES~0299+00 and 1ES~0347-121. The limit is independent on $\lambda_B$ for $\lambda_B \gg D_e$, while it gradually strengthens as $\sqrt{D_e\lambda_B}$ when $\lambda_B\ll D_e$, c.f. Fig.~2 of \cite{neronovscience}. 

Similar results were obtained in \cite{klaus,tavecchio1,tavecchio2}. Such limits have been derived after assuming the blazars to be steady sources of VHE gamma--rays. However, as pointed out in \cite{dermer}, since blazars are known to be strongly variable VHE gamma-ray sources, the assumption of steady flux is tenable only over time scales of the order few years, i.e. the time window over which the blazars have been actually monitored with Cherenkov telescopes and proved to be steady. If such assumption is made, the lower limits on the IGMF strength relax significantly, because the delay time of the cascade emission can easily exceed the observing time window. Thus, a more robust lower limit on the IGMF strength is $B \geq 10^{-18} ... 10^{-17}$~G, also scaling as $1/\sqrt{\lambda_B}$ when $\lambda_B\ll D_e$ \cite{dermer,andrew1,andrew2,takahashi1,takahashi2}. The uncertainty in the above quoted value comes mainly from the uncertainty in the extragalactic background light \cite{andrew2}. 
 

The main aim of this paper is to point out that the dependence of the deflection angle on the IGMF correlation scale $\delta \propto\sqrt{\lambda_B}$ (\Eqref{deltarandom}), adopted in the literature in the limit $\lambda_B\ll D_e$ (see e.g. \cite{neronovsemikoz}), is indeed not always valid. This affects the constraints on the IGMF.
In fact, the square root dependence on $\lambda_B$ follows directly from the assumption that the IGMF is composed of chaotically oriented cells of size $\lambda_B\ll D_e$ inside which the IGMF is correlated. Such a configuration does not satisfy the divergence free condition for the magnetic field. The square root dependence on $\lambda_B$ arises because the charged particle performs a random walk while it propagates on the path $D_e$, since it crosses several uncorrelated cells. 
This picture applies only if the magnetic field spectrum $P_B(k) \propto k^{n_B}$ at large scales (or small wave numbers $k$) satisfies $n_B > -2$. On the other hand, for a very red IGMF spectrum with $-3 < n_B < -2$, the magnetic power is not concentrated on a well defined correlation scale but it continuously grows with scales. Consequently, one cannot assume that the charged particle moves following a random walk. We will show that, for $-3 < n_B < -2$, the dependence of the deflection angle on $\lambda_B$ changes from $\delta \propto \sqrt{\lambda_B}$ to $\delta \propto\lambda_B^{(n_B+3)/2}$. This directly introduces a dependence on the IGMF spectrum in the lower limit that is derived from gamma-ray cascades. In particular, in the limiting case $n_B\ra -3$, the lower limit becomes independent on $\lambda_B$. We will also show that, when the IGMF spectrum is such that $n_B>-2$, even if one goes back to the usual behavior $\delta \propto \sqrt{\lambda_B}$ of  \Eqref{deltarandom}, the dependence on the IGMF spectral index remains, in the form of an overall multiplicative constant that depends on $n_B$.

At scales smaller than the correlation scale $\lambda_B$, it is reasonable to assume that the IGMF spectrum is of the Kolmogorov type. The spectral index of the IGMF at large scales, on the other hand, can directly be connected with the IGMF origin. A magnetic field must be divergence-less, and this property directly implies that the large-scale power spectrum of a {\it causally} created IGMF must be blue: more specifically, $n_B\geq 2$ and even \cite{caprinidurrer1, caprinidurrer2}. This condition must then be satisfied for any IGMF that is generated by astrophysical processes, and also cosmological processes operating during a phase of standard Friedmann expansion of the universe, when the causal horizon is finite: be it today, or during the radiation or matter dominated eras (for a review of cosmological generation processes, see \cite{durrer,kunze}). 

On the other hand, being connected with causality, this constraint on the spectral index does not apply to IGMF that root their origin at inflation. During inflation, in fact, the causal horizon diverges \cite{ruthbook} and magnetic field correlations can be built at larger and lager scales. If the IGMF is generated by inflation, it can therefore be characterized by a red spectral index. In this case, it becomes particularly important to account for the dependence of the lower limit from gamma-ray cascades on the magnetic field spectral index. Note that in this work we consider only non-helical fields, since our aim is only to show how the limits from gamma-ray cascades are modified. For the case of IGMF with non-zero helicity, and the possibility to detect them, see Refs. \cite{tashiro1,tashiro2}. 

The remaining of the paper is organized as follows: in section II, we briefly revise the properties of the cascade initiated by a VHE photon in the IGM; in section III, we define the properties of the stochastic IGMF and evaluate the deflection angle; in section IV we apply the previously derived expression of the deflection angle to the observational bound on the IGMF strength, and discuss in particular the case of a IGMF generated during inflation by the breaking of conformal invariance through the coupling of the electromagnetic field with the background; in section V we conclude. We use CGS units (but in the last section, when discussing the IGMF generated by inflation, we set $c=1$). 

\section{Electromagnetic radiation initiated by a VHE photon}

VHE photons do not travel long distances due to pair-production interactions with soft photons in the extragalactic background light \cite{gould}. Gamma-ray photons of energy $E_{\gamma_0}$ interact mainly with soft photons of wavelength $\lambda_s \simeq 1.5 (E_{\gamma_0}/1~{\rm TeV}) \mu$m (see e.g. \cite{felixbook}). For a blazar located at a distance $D$, the opacity of the pair production process is defined as $\tau = D/D_{\gamma}$, where $D_{\gamma}$ is the mean-free-path of VHE photons. In the multi-TeV domain the following approximate expression can be used \cite{neronovsemikoz}:
\begin{equation}
D_{\gamma} \approx 40 \kappa \left( \frac{E_{\gamma_0}}{20 ~ \rm TeV} \right) ~ {\rm Mpc}
\end{equation}
where for simplicity we neglected the effect of the redshift, and where $\kappa$ is a parameter of order unity that accounts for the uncertainties in the present knowledge of the extragalactic background light (for a review see e.g. \cite{gigi}).

The average energy of secondary electrons (hereafter we refer to both electrons and positrons as electrons) produced in this process is $E_e = E_{\gamma_0}/2$. Electrons of such energy inverse Compton scatter mainly off photons of the cosmic microwave background radiation, of typical energy $\epsilon_{\rm CMB} = 3 k_B T_{\rm CMB} \sim 7 \times 10^{-4}$~eV, where $k_B$ is the Boltzmann constant and $T_{\rm CMB}$ the temperature of the photon field. 
The cooling length of such electrons is (e.g. \cite{felixbook}):
\begin{equation}
D_e = \frac{3 m^2 c^4}{4 \sigma_T E_e \omega_{\rm CMB}} \simeq 40 \left( \frac{E_{\gamma_0}}{20~\rm TeV} \right)^{-1} \rm kpc
\end{equation}
while the energy of the up-scattered photons reads (e.g. \cite{felixbook}):
\begin{equation}
E_{\gamma} = \frac{4}{3} \left( \frac{E_e}{m c^2} \right)^2 \epsilon_{\rm CMB} \simeq 0.4 \left( \frac{E_{\gamma_0}}{20~{\rm TeV}} \right)^2 \rm TeV
\end{equation}
where $\omega_{\rm CMB} \simeq 0.25$~eV/cm$^3$ is the energy density of the cosmic microwave background radiation, $\sigma_T$ the Thompson cross section, and $m c^2$ the electron rest mass energy.
Thus, the secondary gamma-ray emission has to be searched in the sub-TeV energy domain.

\section{Stochastic IGMF and deflection angle}

We model the IGMF as a statistically homogeneous and isotropic random field with $\vev{B_i(\bx)}=0$, spatial correlation function
\be
\vev{B_i(\bx)B_j(\bx+\by)}=\xi_{ij}(|\by|) \label{corrfun}
\ee
and energy density corresponding to the squared rms strength: $\rho_B=\vev{B^2}/(8 \pi)=\xi_{ii}(0)/(8 \pi)$. In Fourier space one has then \cite{durrer}
\bea
&\vev{B_i(\bk)B_j^*(\bq)}=(2\pi)^3 \mathcal{P}_{ij}(k) \de(\bk-\bq)\,, \label{BkBq}\\
&\mathcal{P}_{ij}(k) = \int d^3 y \,{\rm e}^{i \bk\cdot\by} \xi_{ij}(|\by|) \equiv \frac{1}{2}(\de_{ij}-\hat k_i\hat k_j) P_B(k). \label{Pij} 
\eea
The projector after the second equality in the second line follows from the requirement that ${\rm div} \,{\bf B}=0$ (see e.g. \cite{caprinidurrer2}). Note that we only consider non-helical IGMF, as motivated in the introduction. The IGMF power spectrum $P_B(k)$ can be modeled with two power laws:
\bea
P_B(k)=A \left\{\begin{array}{ll} \vspace*{0.2cm}\big(\frac{k}{k_0}\big)^{n_B} & {\rm if}~k\leq k_0 \\
\big(\frac{k_0}{k}\big)^{m_B} & {\rm if}~k> k_0
\end{array}\right.
\label{PBk}
\eea
where we have introduced a characteristic wavenumber $k_0$ at which the spectrum changes slope. In order to make the comparison with previous analyses, we define the magnetic field correlation scale as 
\be
\lambda_B\equiv \frac{2\pi}{k_0}\,.
\ee 
Therefore, we assume that this is the physical parameter to be eventually constrained by the data, together with the IGMF rms strength $\sqrt{\vev{B^2}}$. Consequently, the normalisation constant $A$ is better rewritten in terms of $\sqrt{\vev{B^2}}$: using Eqs.~\eqref{BkBq}, \eqref{Pij} one finds
\be
\rho_B=\frac{\vev{B^2}}{8 \pi}=\frac{1}{2(2\pi)^3}\int_0^\infty dk\,k^2\,P_B(k)\,, \label{rhoB}
\ee
which implies then, with \Eqref{PBk}
\be
A=\frac{(2\pi)^2}{2}\frac{(n_B+3)(m_B-3)}{n_B+m_B}\frac{\vev{B^2}}{k_0^3}\,.
\ee
For the energy density not to diverge, the spectral indexes must satisfy the conditions $n_B>-3$, $m_B>3$. At small scales $k>k_0$, the IGMF has been processed by MHD turbulence during its evolution from its generation until today: it is therefore natural to assume $m_B=11/3$, corresponding to Kolmorogorov turbulence. On the other hand, at large scales $k<k_0$ the IGMF has not been modified by the turbulent cascade: the spectral index $n_B$ is directly related to the process that generated the IGMF and it is one of the relevant parameters characterizing the IGMF (see \cite{durrer} and references therein). 

The characteristic scale which is usually assigned to the stochastic magnetic field as ``correlation" scale is the integral scale \cite{durrer}
\be
\lambda_I\equiv \frac{1}{8\pi^2}\frac{1}{\rho_B} \int_0^\infty dk\,k\,P_B(k)\,.
\ee
From \Eqref{PBk} it appears that this quantity is only well defined for $n_B>-2$. In this case, $\lambda_I$ is the scale at which most of the magnetic energy is concentrated, and it is directly related to the parameter we defined as the correlation scale $\lambda_B=2\pi/k_0$: 
\be
\lambda_I= \lambda_B \,\frac{2(n_B+3)}{5(n_B+2)}~~~~{\rm for}~n_B>-2\,.
\ee
Therefore, when $n_B>-2$ it is natural to assign to the IGMF a correlation scale which is simply $\lambda_B\simeq \lambda_I$. 

However, the only condition one has to impose on $n_B$ is $n_B>-3$ for the IGMF energy density not to diverge. Therefore it is (at least in principle) possible to have IGMF with very red large scale power spectra, $-3<n_B\leq -2$. These cannot have a causal origin, but can be produced during inflation \cite{durrer,kunze}. In this case, $k_0=2\pi/\lambda_B$ is not a correlation scale in the same sense as $\lambda_I$, but it corresponds to the upper cutoff of the IGMF spectrum at generation time, which is directly related to the process of generation: in general, for an inflationary IGMF, it is given by the inverse Hubble scale during inflation $k_0\simeq H_{\rm inf}$. In the following, we keep the definition $\lambda_B=2\pi/k_0$ to be able to describe both cases ($n_B>-2$ and $-3<n_B\leq -2$) with only one parameter. We now proceed to demonstrate that, if one accounts for a generic large scale spectral index $n_B$, the limit from the gamma-ray cascades on the IGMF strength must be modified with respect to those presented so far in the literature.  

In order to derive a limit on the IGMF strength from gamma-ray observations we need to compute the deflection angle $\delta$ of the electrons produced in pair-production interactions of VHE photons \cite{plaga,neronovsemikoz}.
In order to evaluate the dependence of the deflection angle on the statistical properties of the IGMF we follow Ref.~\cite{aharonian} (for a more refined analysis accounting for the three-dimensional deviation of the charged particle, see Ref.~\cite{tashiro1}). The basic assumption is that the deflection experienced by the electron in the IGMF is in any case small. Over distances smaller than $D_e$ the electron energy $E_e$ remains approximately constant, and therefore its motion in the IGMF is described by the equation $\dot{\bf v}=(e\,c/E_e) [{\bf v}\times{\bf B}({\bf r})]$, where ${\bf r}$ denotes the position of the particle. The particle is ultra-relativistic with ${\bf v}=c \,\hat{{\bf n}}$, and the variation of the velocity is $\dot{\bf v} \perp {\bf v}$. Therefore, one can calculate the deflection angle over a time interval $\Delta t$ as~\cite{aharonian}
\be
\Delta {\bf v}= \delta \,c\, \hat{\bf n}_\perp = \frac{e\,c^2}{E_e}\int_t^{t+\Delta t} [\hat{{\bf n}}\times {\bf B}({\bf r})]\, dt\,.
\ee
Assuming that the particle propagates in the $z \parallel \hat{\bf n}$ direction almost unchanged by the presence of the IGMF, the rms deflection can be expressed as ($dz=c\,dt$)
\bea
&\vev{\de^2}= \left(\frac{e}{E_e}\right)^2 \times&\\  
&\int_z^{z+\Delta z} dz_1 \int_z^{z+\Delta z} dz_2 \, \vev{[\hat{{\bf n}}\times {\bf B}({\bf r}_1)]\cdot[\hat{{\bf n}}\times {\bf B}({\bf r}_2)]} \simeq & \nonumber \\
& \left(\frac{e}{E_e}\right)^2 [\de_{ij}-\hat n_i \hat n_j]   \int_z^{z+\Delta z} dz_1 \int_z^{z+\Delta z} dz_2 \,\vev{B_i({\bf r}_1) B_j({\bf r}_2)}\,. & \nonumber
\eea
\Eqref{corrfun} together with the assumption of small deflection $|{\bf r}_1-{\bf r}_2|\simeq |z_1-z_2|\equiv \zeta$ gives 
\be
\vev{\de^2}\simeq 2\, \Delta z \left(\frac{e}{E_e}\right)^2 [\de_{ij}-\hat n_i \hat n_j]  \int_0^{\Delta z} d\zeta\, \xi_{ij}(\zeta)\,.
\ee
This can be calculated in terms of the IGMF power spectrum substituting the inverse Fourier transform of \Eqref{Pij} and performing the integral on $\zeta$ and on the angle $d\Omega_{\hat{k}}$ (where $\int d^3 k =\int_0^\infty dk\,k^2 \int_{4\pi} d\Omega_{\hat{k}} $), to arrive at 
\bea
\vev{\de^2}&\simeq& \frac{2}{(2\pi)^2}\left(\frac{e}{E_e}\right)^2 D_e \int_0^\infty dk\,k\,P_B(k) \times \\
&&\left[ \frac{\sin{kD_e}-kD_e\cos{kD_e}+(kD_e)^2{\rm Si}(kD_e)}{(kD_e)^2}\right]\, \nonumber
\eea
where we set $\Delta z=D_e$ to obtain the total deflection suffered by an electron over the cooling time, and ${\rm Si}(kD_e)$ denotes the sinus integral function \cite{abramovitz}. The above integral can be performed analytically. With $R_L=\sqrt{\vev{B^2}}E_e/e$ one arrives at a simple expression for the rms deflection angle:
\be
\sqrt{\vev{\de^2}}\simeq \frac{\sqrt{D_e\,\lambda_B}}{R_L}~ \Pi \left(\frac{D_e}{\lambda_B},n_B\right)\,, \label{rmsde}
\ee
where $\Pi \left(\frac{D_e}{\lambda_B},n_B\right)$ is an analytic function (given more explicitly in the appendix), plotted in Fig.~\ref{fig:Pi} for different values of $n_B$.
As expected, in the limit $\lambda_B\gg D_e$, the dependence on $n_B$ disappears, while if $\lambda_B\ll D_e$ the function $\Pi\left(D_e/\lambda_B,n_B\right)$ varies with the IGMF large-scale spectral index $n_B$. The asymptotic behavior of $\Pi \left(\frac{D_e}{\lambda_B},n_B\right)$ is such that
\begin{widetext}
\bea
\sqrt{\vev{\de^2}}\simeq \frac{\sqrt{D_e\,\lambda_B}}{R_L}
\left\{\begin{array}{ll} 
\frac{2}{\sqrt{3}} \sqrt{\frac{D_e}{\lambda_B}} +\mathcal{O}\left(\frac{D_e}{\lambda_B}\right)^{7/6} & {\rm if}~\lambda_B\gg D_e\,, \\
\left[ \left( \frac{\lambda_B}{D_e} \right)^{n_B+2} 
\left( \frac{(n_B+1)\Gamma(n_B+4)\sin(n_B\pi/2)}{\pi^2(2\pi)^{n_B+1}n_B(n_B+2)^2(3n_B+11)} \right) 
+\frac{n_B+3}{10\,n_B+20}+\mathcal{O}\left(\frac{\lambda_B}{D_e}\right)^{3} \right]^{\frac{1}{2}} & {\rm if}~\lambda_B\ll D_e\,.
\end{array}\right. 
\label{delimit}
\eea
\end{widetext}
From the upper row of \Eqref{delimit} one sees that, if the correlation scale $\lambda_B$ is very large, the deflection angle goes back to the usual expression \Eqref{deltahom}: it is as if the particle would propagate in a constant IGMF. The small numerical factor $2/\sqrt{3}$ is due to the fact that we calculate the deflection angle a bit more precisely as an integral over the charged particle path, while \Eqref{deltahom} is approximate. On the other hand, when the correlation scale is small\footnote{Note that this is usually the case for inflationary generated IGMF, for which the upper cutoff $k_0\simeq H_{\rm inf}$ corresponds to very small scales today, even after evolution by MHD processing.}, one does not recover \Eqref{deltarandom}: an explicit dependence on $n_B$ appears in the deflection angle. If $n_B> -2$, the asymptotic behavior given in the second row of \Eqref{delimit} is such that the expression in brackets becomes independent on $\lambda_B$: the constant term dominates. Therefore, the deflection angle goes back to the usual expression $\sqrt{\vev{\de^2}}\propto \sqrt{D_e\lambda_B}/R_L$, but it still differs from \Eqref{deltarandom} by a multiplicative constant: $\sqrt{(n_B+3)/(10\,n_B+20)}$. For sufficiently red spectra $-3<n_B<-2$, instead, the first term in brackets of the second row of \Eqref{delimit}  dominates over the constant. The dependence of the deflection angle on $\lambda_B$ changes with respect to the one given in \Eqref{deltarandom}, and becomes explicitly dependent on $n_B$: $\sqrt{\vev{\de^2}}\propto \lambda_B^{(n_B+3)/2}$. Note that the divergence for $n_B=-2$ in the analytic expressions of \Eqref{delimit} is only apparent. 

\begin{figure}
\includegraphics[scale=0.85]{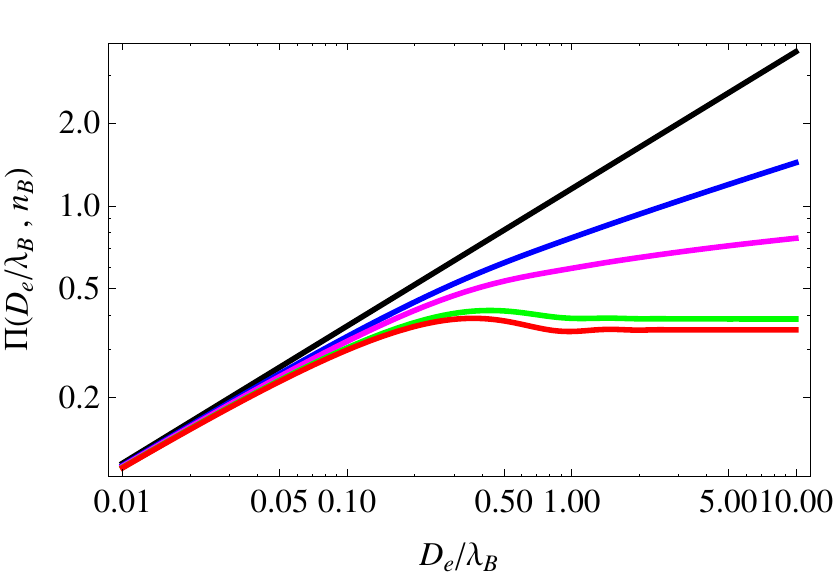}
\caption{The function $\Pi \left(\frac{D_e}{\lambda_B},n_B\right)$ defined in \Eqref{rmsde} for different values of the IGMF large scale spectral index: from top to bottom $n_B\rightarrow -3$ (black), $n_B=-2.5$ (blue), $n_B= -2$ (magenta), $n_B=0$ (green), $n_B=2$ (red). For large vs. small correlation scale, the function reaches the asymptotic behaviors given in \Eqref{delimit}.}
\label{fig:Pi}
\end{figure}

\begin{figure}
\includegraphics[scale=0.85]{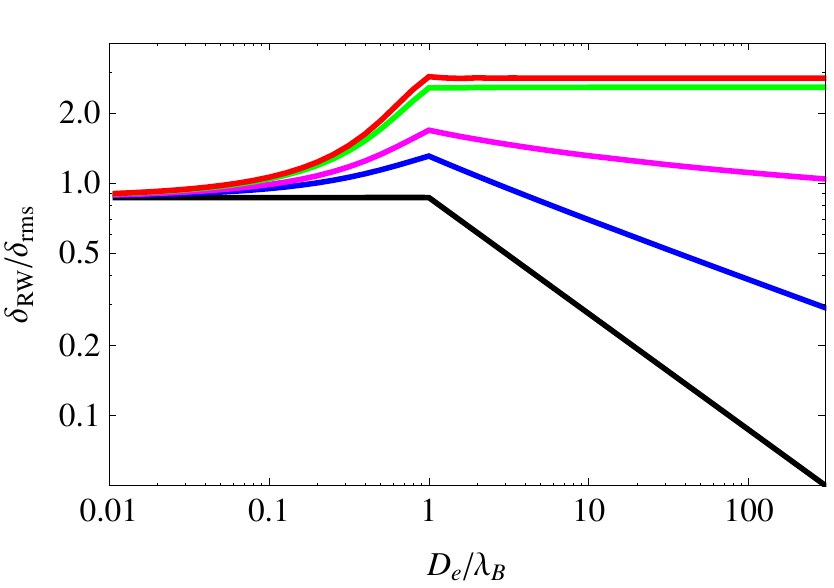}
\caption{The ratio between the deflection angle taken from the literature $\delta_{\rm RW}$ (Eqs.~\eqref{deltarandom} and \eqref{deltahom}) the one evaluated here,  \Eqref{deoverde}. From bottom to top, $n_B\rightarrow -3$ (black), $n_B=-2.5$ (blue), $n_B= -2$ (magenta), $n_B=0$ (green), $n_B=2$ (red). The two angles are the same for large correlation scale $\lambda_B$ (a part from a small numerical factor), but diverge if the correlation scale is small, depending on the value of $n_B$. If $n_B>-2$, they only differ by a multiplicative constant that depends on $n_B$ (see main text). The spikes in the figure are due to the fact that $\de_{\rm RW}$ is only defined in the limiting cases $\lambda_B\gg D_e$ and $\lambda_B\ll D_e$, and we join them at $\lambda_B=D_e$.}
\label{fig:deRW}
\end{figure}

This behavior is also seen by comparing directly \Eqref{rmsde} with the usual deflection angle given in the literature (Eqs.~\eqref{deltarandom} and \eqref{deltahom}), based on the random walk model and therefore here denoted $\delta_{\rm RW}$:
\bea
\label{deoverde}
\frac{\delta_{\rm RW}}{\sqrt{\vev{\de^2}}}\simeq \Pi^{-1} \left(\frac{D_e}{\lambda_B},n_B\right)\left\{\begin{array}{ll} \vspace*{0.2cm}
\sqrt{\frac{D_e}{\lambda_B}} & {\rm if}~\lambda_B\gg D_e\,, \\
1 & {\rm if}~\lambda_B\ll D_e\,.
\end{array}\right. 
\eea
This quantity is plotted in Fig.~\ref{fig:deRW}.

\section{IGMF constraints from gamma-ray cascades}

The lower limit on the IGMF strength that has been obtained from gamma-ray observations gets modified correspondingly to the change in the deflection angle derived above. In order to illustrate this effect, we consider the lower limit from the observations of delayed emission which are derived in \cite{durrer}: in particular, we reproduce the result shown in Fig.~14 of this reference. 

\begin{figure}
\includegraphics[scale=0.98]{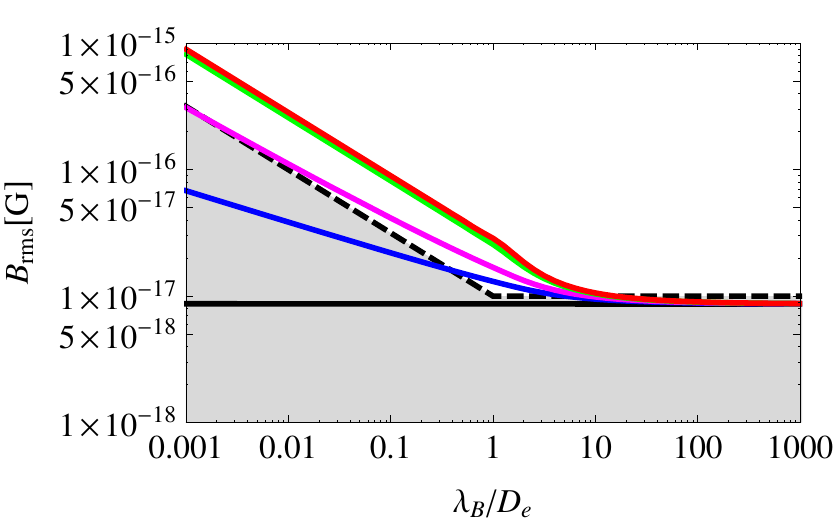}
\caption{Lower bound on the IGMF strength as a function of $\lambda_B/D_e$. The grey-shaded region below the dashed curve corresponds to the excluded region of Fig.~14 of \cite{durrer}. The other curves are the lower bounds on $\sqrt{\vev{B^2}}$ for, from bottom to top, $n_B\rightarrow -3$ (black), $n_B=-2.5$ (blue), $n_B= -2$ (magenta), $n_B=0$ (green), $n_B=2$ (red). }
\label{fig:Blim}
\end{figure}

Neglecting redshift dependences, the expression of the time delay becomes \cite{neronovsemikoz}:
\bea
t_{\rm delay}&\simeq& \frac{D_\gamma}{2 c} \,\vev{\delta^2} \\
&\simeq& \frac{D_\gamma}{2 c} \left(\frac{D_e}{R_L}\right)^2 \frac{\lambda_B}{D_e}\,\Pi^2\left(\frac{D_e}{\lambda_B},n_B\right)\,, \nonumber
\eea
where in the second line we have substituted \Eqref{rmsde}.
In the case $\lambda_B\gg D_e$, when $\de_{\rm RW}=D_e/R_L$, we can directly compare the above expression to the one given in Eq.~(151) of \cite{durrer}, to find
\bea
&t_{\rm delay} \simeq & 0.3\, \left[\frac{E_\gamma}{0.1\,{\rm TeV}}\right]^{-\frac{5}{2}} \frac{\vev{B^2}}{(10^{-17}\,{\rm G})^2} \times \nonumber \\
 &&  \frac{\lambda_B}{D_e}\,\Pi^2\left(\frac{D_e}{\lambda_B},n_B\right) {\rm yr}~~~~{\rm if}~\lambda_B\gg D_e\,, 
\eea
which, choosing $E_\gamma=0.1$ TeV and $t_{\rm delay}=0.3$ yr, directly translates into the limit
\be
\sqrt{\vev{B^2}}_{\rm lim} \simeq 10^{-17}\,{\rm G}\,\, \sqrt{\frac{D_e}{\lambda_B}}\,\,\Pi^{-1}\left(\frac{D_e}{\lambda_B},n_B\right)\,. \label{thebound}
\ee
This limit is shown in Fig.~\ref{fig:Blim} together with the one from Fig.~14 of \cite{durrer}, for several values of $n_B$. For large correlation scale, our limit is comparable to the one given in \cite{durrer} (a part from the negligible small numerical factor given in the upper line of \Eqref{delimit}). For small correlation scale instead, our limit depends on $n_B$. In particular, if $n_B> -2$ the dependence on $\lambda_B$ is the same as the usual one $\sqrt{\vev{B^2}}_{\rm lim}\propto \sqrt{D_e/\lambda_B}$, but our limit is more constraining than the one of \cite{durrer} by a multiplicative constant that depends on $n_B$. This is due to the particular shape of the function $\Pi$: from \Eqref{delimit}, one  infers the constant
\bea
&&\sqrt{\vev{B^2}}_{\rm lim} \sim 10^{-17}\,{\rm G}\,\, \sqrt{\frac{D_e}{\lambda_B}}\,\, \sqrt{\frac{10\,n_B+20}{n_B+3}} \nonumber\\
&&~~~~~~~~~~~~~~~~~~~~~~ {\rm for}~~~\lambda_B\ll D_e\,,~n_B> -2\,. \label{constant}
\eea
When $-3<n_B< -2$, instead, our limit is relaxed with respect to the one of Ref.~\cite{durrer}. In particular, it becomes independent on $\lambda_B$ for a scale invariant IGMF (c.f. again \Eqref{delimit}). 

As mentioned in the introduction, IGMF generated by a causal process (anything a part from inflation) have blue spectra such that $n_B\geq 2$ and even \cite{caprinidurrer1, caprinidurrer2}. In this case, we have seen that the usual lower bound from gamma-ray cascades should get modified only by a multiplicative constant depending on $n_B$, but it retains the usual dependence on $\lambda_B$ as $\sqrt{\vev{B^2}}_{\rm lim}\propto \sqrt{D_e/\lambda_B}$. 

\begin{figure}
\includegraphics[scale=0.23]{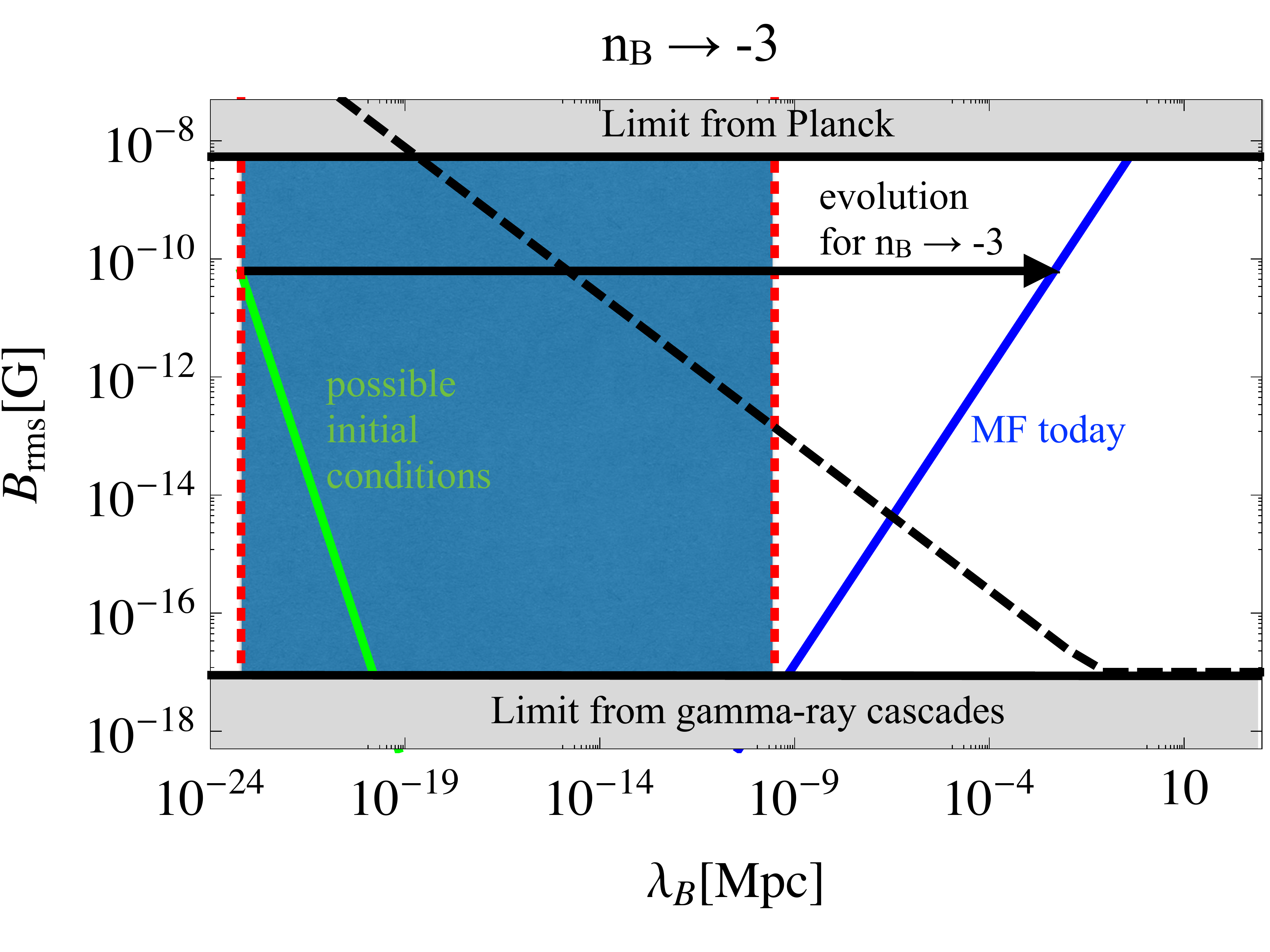}
\caption{The blue-shaded area corresponds to the observationally testable region for inflationary generated magnetic fields in scenarios with inflaton coupling to the electromagnetic field. We show here the case $n_B\rightarrow-3$. The upper and lower gray-shaded regions are excluded respectively by the Planck upper bound on the intensity of a primordial magnetic field \cite{planck} and by the bound given in \Eqref{thebound}. The black, dashed line is the bound given in Ref.~\cite{durrer}. The vertical red dashed lines correspond to the minimal and maximal correlation scale, depending on the energy scale of inflation (c.f. main text). The green line represents the set of possible initial conditions given in Eqs.~\eqref{vevBinf} and \eqref{corrBinf} with $\alpha_1=\alpha_2=1$. The black, horizontal arrow represents one example of the evolution of the IGMF through freely-decaying MHD. The endpoints of the evolution must fall on the blue line given by \Eqref{final}. We have taken for the energy loss length of the electron $D_e\simeq 80$ kpc, corresponding to the observed photon energy $E_\gamma=0.1$ TeV (see section II). }
\label{fig:Binfmen3}
\end{figure}

\begin{figure}
\includegraphics[scale=0.23]{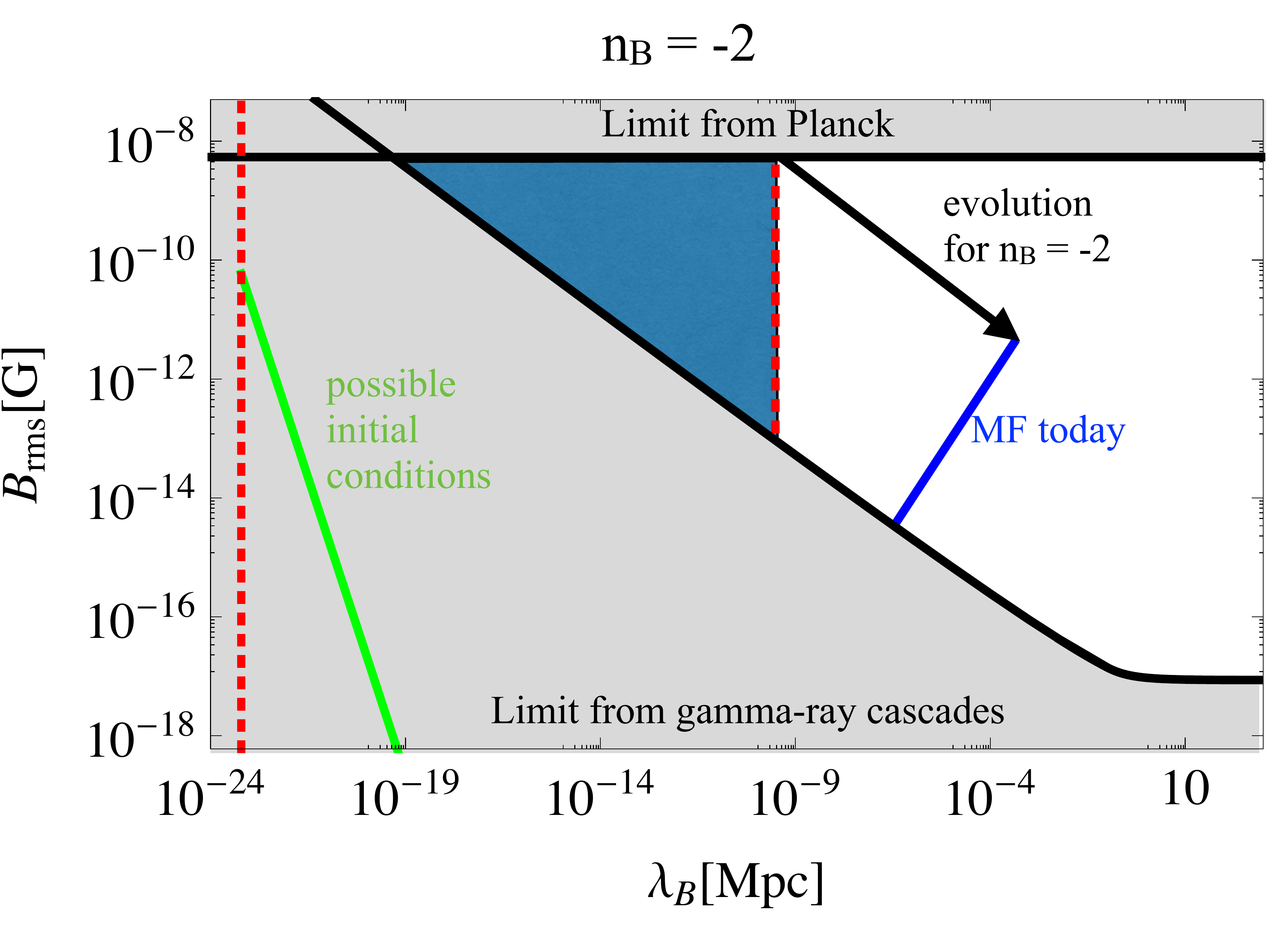}
\caption{The same as Fig.~\ref{fig:Binfmen3} but for $n_B=-2$.}
\label{fig:Binfmen2}
\end{figure}

IGMF with spectra $n_B<2$ could only be generated by inflation. For a review of inflationary generation mechanisms, see Refs.~\cite{durrer,kunze}. In order to illustrate the effect of the new IGMF lower bound found above, we reproduce here Fig.~16 of~\cite{durrer}, which shows the observationally testable region for inflationary IGMF in scenarios where the IGMF generation is based on the breaking of conformal invariance through the coupling of the electromagnetic field with the background (via direct coupling to the inflaton, or to another scalar field). We reproduce the figure for two values of the spectral index: $n_B\rightarrow -3$ shown in Fig.~\ref{fig:Binfmen3}, and $n_B=-2$ shown in Fig.~\ref{fig:Binfmen2}. 

The outcome of the inflationary generation process, i.e. the IGMF amplitude, correlation scale, power spectrum, depend on the details of generation mechanism. However, for scenarios based on the coupling of the electromagnetic field to the background, one has in general that the rms IGMF amplitude and its correlation scale, redshifted to today, are given by
\bea
\sqrt{\vev{B^2}}_{\rm reh}&\simeq& \alpha_1 \,H_{\rm inf}^2 \left(\frac{T_0}{T_{\rm reh}}\right)^2 \left(\frac{g_0}{g_{\rm reh}}\right)^{2/3}\, \label{vevBinf} \\ 
\lambda_B^{\rm reh}&\simeq& \frac{\alpha_2}{H_{\rm inf}}  \left(\frac{T_{\rm reh}}{T_0}\right) \left(\frac{g_{\rm reh}}{g_0}\right)^{1/3} \label{corrBinf} \,.
\eea
Here we have inserted the multiplicative factors $\alpha_1$ and $\alpha_2$ whose amplitude depends on the particular generation process. Moreover, $H_{\rm inf}$ is the Hubble scale at inflation, $T_0$ and $T_{\rm reh}$ are the temperatures today and at reheating time, and correspondingly $g_0=3.9$ and $g_{\rm reh}=106.75$ are the number of relativistic degrees of freedom. The Hubble scale is such that $3H_{\rm inf}^2M_{\rm P}^2\simeq \rho_{\rm reh}=a_{\rm sb}g_{\rm reh}T_{\rm reh}^4$, with $M_{\rm P}$ the reduced Planck mass and $a_{\rm sb}$ the Stefan-Boltzmann constant. The initial conditions given in Eqs.~\eqref{vevBinf} and \eqref{corrBinf} are shown by the green line in Figs.~\ref{fig:Binfmen3} and \ref{fig:Binfmen2}, from the maximal reheating temperature of $T_{\rm reh}\simeq 10^{16}$ GeV down to the minimal one, which we set at $T_{\rm reh}\simeq 200$ GeV as in \cite{durrer}. To plot $\sqrt{\vev{B^2}}_{\rm reh}$ as a function of $\lambda_B^{\rm reh}$, we have set $\alpha_1$ and $\alpha_2$ to one in Eqs.~\eqref{vevBinf} and \eqref{corrBinf}: this is why we stress that these are only ``possible" initial conditions. 

The maximal and minimal reheating temperature also correspond to a minimal and maximal correlation scale, following \Eqref{corrBinf}: setting $\alpha_2=1$, one has $\lambda_B^{\rm reh}(10^{16}\,{\rm GeV})\simeq 6\times 10^{-24}$ Mpc and $\lambda_B^{\rm reh}(200\,{\rm GeV})\simeq 3\times 10^{-10}$ Mpc. These are represented by the left, respectively right red-dashed vertical lines in Figs.~\ref{fig:Binfmen3} and \ref{fig:Binfmen2}. 

After generation, the IGMF evolves during the expansion of the universe according to free turbulent decay. For a non-helical magnetic field, the evolution is such that the final values of the IGMF and its correlation scale evolve through selective decay, giving \cite{caprinidurrer3,durrer}
\be
\sqrt{\vev{B^2}} \, \lambda_B^{\frac{n_B+3}{2}}= {\rm const} = \sqrt{\vev{B^2}}_{\rm in} \, ({\lambda_B^{\rm in}})^{\frac{n_B+3}{2}}\,,
\label{evol}
\ee
with $\sqrt{\vev{B^2}}_{\rm in}$ and  ${\lambda_B^{\rm in}}$ the initial conditions. Note that, if $n_B=-2$, the selective decay evolution is the same as the one of an helical magnetic field (inverse cascade due to conservation of helicity). Examples of the selective decay evolution are shown by the black arrows in Figs.~\ref{fig:Binfmen3} and \ref{fig:Binfmen2}. The endpoints of the evolution by free turbulent decay are shown by the blue solid line, and must be such as to satisfy the relation \cite{banerjee,durrer}
\be
\sqrt{\vev{B^2}} = 10^{-8} \, \frac{\lambda_B}{\rm Mpc} \, {\rm G}\,.
\label{final}
\ee

The observationally testable region for inflationary IGMF is given by the set of all possible initial conditions, i.e. couples of values $(\sqrt{\vev{B^2}}_{\rm in},\lambda_B^{\rm in})$ which, following the evolution given in \Eqref{evol}, can fall on the end-line \eqref{final}. With respect to the result given in Fig.~16 of~\cite{durrer}, here we have that in the case $n_B\rightarrow -3$ the testable region opens up to include also the ``possible" initial conditions given in Eqs.~\eqref{vevBinf} and \eqref{corrBinf}: c.f. the blue-shaded area in Fig.~\ref{fig:Binfmen3}. For $n_B=-2$, on the other hand, these initial conditions remain excluded: they are not contained in the blue-shaded area of Fig.~\ref{fig:Binfmen2}. For intermediate values of $-3<n_B<-2$, the testable region would change as shown in Fig.~\ref{fig:Blim}. For $n_B>-2$, the testable region differs from the one shown in Fig.~16 of~\cite{durrer} only by the effect of the overall constant (c.f \Eqref{constant}): the lower limit on $\sqrt{\vev{B^2}}$ becomes a bit more restrictive, and the observationally testable region shrinks accordingly. Consequently, testable IGMF with $n_B\geq -2$ can be produced only if the values of $\alpha_1$ and $\alpha_2$ are significantly large than one. This can actually happen quite easily: see e.g. the model studied in Ref.~\cite{Caprini:2014mja} and references therein. Another possibility is, if the magnetic field gets strongly amplified by dynamo action during reheating. The same conclusion had been drawn in Ref.~\cite{durrer}. There are currently no studied mechanisms that can lead to a large amplification during reheating, but this is not excluded. For example, recently Ref.~\cite{Fujita:2015iga} has studied in details the post-inflationary evolution of helical magnetic fields generated by a pseudoscalar coupling to the inflaton in a $m^2\phi^2$ inflationary model. 

To summarize, inflationary scenarios that would provide IGMF with spectral indexes $n_B\geq -2$ are more constrained than previously thought, since the observationally testable region becomes more restrictive than what derived in \cite{durrer}. On the other hand, scenarios that would provide $-3<n_B<-2$ are less constrained. Note, however, that valid IGMF generation mechanisms based on the breaking of conformal invariance through the coupling of the electromagnetic field with the inflaton always tend to give IGMF with blue spectra with $n_B\geq 1$ \cite{durrer}. The most widely studied generation mechanism which could provide $n_B\rightarrow -3$, first proposed in \cite{turner}, has a strong-coupling problem \cite{demozzi}. Still, we find it useful to present how the constraints from gamma-ray cascades would change, in case in the future some new, well-behaved IGMF generation mechanism is proposed, that can lead to red spectral indexes $-3 < n_B < -2$.

\section{Conclusions}

The non-observation of cascade radiation from blazars has been used to put constraints on the strength of the IGMF. The IGMF can be described as a stochastic field, fully characterized by its rms strength, correlation scale, and power-law power spectrum. One expects therefore that the constraints inferred from the blazar observations depend on all these parameters. Indeed, it as been found that the lower bound on the IGMF rms strength depends on the correlation scale in such a way that it strengthens as $\sqrt{\vev{B^2}_{\rm lim}}\propto \sqrt{D_e/\lambda_B}$ for small correlation scales $\lambda_B\ll D_e$, while it is independent of $\lambda_B$ in the opposite regime. However, no dependence on the IGMF power spectrum had been identified in previous analyses. The inverse square root dependence on $\lambda_B$ comes directly from the assumption that the IGMF is composed of chaotically oriented cells of size $\lambda_B\ll D_e$, inside which the IGMF is constant and therefore correlated. Here we have used a more refined model to evaluate the deflection angle experienced by the electron traveling through the IGMF, a model that accounts for the properties of the IGMF power spectrum. We have found that in general the lower limit inferred by blazar observations depends also explicitly on the IGMF power spectrum: in particular, on its large scale spectral index $n_B$. If $n_B> -2$, the above mentioned cell-model does effectively describe the real situation: in this case, we recover in fact the behavior $\sqrt{\vev{B^2}_{\rm lim}}\propto \sqrt{D_e/\lambda_B}$ when $\lambda_B\ll D_e$. However, we have demonstrated that an overall multiplicative constant depending on $n_B$ must be included in the amplitude, which strengthens the constraint somewhat (c.f. Eq.~\eqref{constant}). If instead $-3<n_B<-2$, the dependence on the correlation scale when $\lambda_B\ll D_e$ is changed to $\sqrt{\vev{B^2}_{\rm lim}}\propto (D_e/\lambda_B)^{(n_B+3)/2}$. This means that, if the IGMF power spectrum is sufficiently red, one cannot use the cell-model since $\lambda_B$ is not the typical scale at which most the IGMF energy density is concentrated. 

In general, we can conclude that the constraints inferred from the blazar observations must be modified as given in \Eqref{thebound}: i.e., denoting $B_{\rm RW}$ the upper limit from previous analyses (in the regime in which it is constant, $\lambda_B\gg D_e$), one has 
\be
\sqrt{\vev{B^2}}_{\rm lim} \simeq B_{\rm RW}\,\, \sqrt{\frac{D_e}{\lambda_B}}\,\,\Pi^{-1}\left(\frac{D_e}{\lambda_B},n_B\right)\,, 
\ee
where the function $\Pi\left(\frac{D_e}{\lambda_B},n_B\right)$ (given in the appendix) is plotted in Fig.~\ref{fig:Pi}. This equation represents the main result of this paper. Note that in \Eqref{thebound} we have taken as an example $B_{\rm RW}=10^{-17}\,{\rm G}$: this value reproduces the constraint given in paragraph 5.6.3 and Fig.~16 of \cite{durrer}. 

For the most relevant scenarios of IGMF generation, the usual dependence $\sqrt{\vev{B^2}_{\rm lim}}\propto \sqrt{D_e/\lambda_B}$ continues to hold, and the lower bounds from blazar observations depend mildly on $n_B$: only in the amplitude and not in the $\lambda_B$-slope. Causally generated IGMF, in fact, are all characterized by $n_B\geq 2$ and even. Very negative spectral indexes $-3<n_B<-2$ can only arise if the IGMF has been produced during inflation, and no completely consistent inflationary model has been identified yet which does lead to such red spectra. However, in the future some new, well-behaved inflationary generation mechanism may be proposed, that can lead to IGMF with red spectra. In this case, the present result implies that the initial conditions it would have to provide to fulfill the lower bounds from gamma-ray cascades are significantly less constrained than previously thought. \\

\appendix*
\section{}

The function $\Pi\left(\frac{D_e}{\lambda_B},n_B\right)$ is given by 
\bea
&\Pi\left(\frac{D_e}{\lambda_B},n_B\right) &= \sqrt{\frac{n_B+3}{\pi(3n_B+11)}} \times \\
&&\left[  \left( \frac{2\pi\,D_e}{\lambda_B}\right)^{n_B+2} \int_0^{\frac{2\pi\,D_e}{\lambda_B}}
dx \,x^{n_B-1} f(x)\right.  \nonumber\\ 
&&+ \left.\left( \frac{2\pi\,D_e}{\lambda_B}\right)^{5/3} \int_{\frac{2\pi\,D_e}{\lambda_B}}^\infty dx \,x^{-14/3} f(x) \right]^{1/2} \nonumber
\eea
with 
\be
f(x)=\sin(x)-x\cos(x)+x^2\,{\rm Si}(x)\,.
\ee
This can be integrated analytically.

\end{document}